\documentclass[11pt]{article}
\normalsize 
\usepackage[margin=1in]{geometry} 
\usepackage{mathtools}

\begin{document}

\title{Error Tree: A Tree Structure for Hamming \& Edit Distances \& Wildcards Matching}

\author{Anas Al-Okaily \\
Computer Science \& Engineering Department, University of Connecticut\\
aaa10013@engr.uconn.edu}
\date{}
\maketitle

\abstract {Error Tree is a novel tree structure that is mainly oriented to solve the approximate pattern matching problems, Hamming and edit distances, as well as the wildcards matching problem. The input is a text of length $n$ over a fixed alphabet of length $\Sigma$, a pattern of length $m$, and $k$. The output is to find all positions that have $\leq$ $k$ Hamming distance, edit distance, or wildcards matching with $P$. The algorithm proposes for Hamming distance and wildcards matching a tree structure that needs $O(n\frac{log_\Sigma ^{k}n}{k!})$ words and takes $O(\frac {m^k}{k!} + occ$)($O(m + \frac {log_\Sigma ^kn}{k!} + occ$) in the average case) of query time for any online/offline pattern, where $occ$ is the number of outputs. As well, a tree structure of $O(2^{k}n\frac{log_\Sigma ^{k}n}{k!})$ words and $O(\frac {m^k}{k!} + 3^{k}occ$)($O(m + \frac {log_\Sigma ^kn}{k!} + 3^{k}occ$) in the average case) query time for edit distance for any online/offline pattern.

\section{Introduction}
In the middle of the increasing growth of the internet-based searching, information retrieval, data mining applications, and bioinformatics researches; there is an increasing necessity of the problem of searching whether a given pattern happened to occur as exact match, approximate match, or wildcards match in a given database. The pattern is usually of small size such as words or sentences, and the database is of much larger size such as web documents, genomes, and books.

The exact matching problem is the simplest form of the pattern matching problems, while the approximate and wildcards matching are more complicated. For the exact matching problem, \cite{weiner1973linear} proposed a tree structure which was improved by \cite{mccreight1976space} and \cite{DBLP:journals/algorithmica/Ukkonen95} and led to an optimal solution of linear tree structure and linear query time. 

Approximate matching involves two problems; Hamming distance and edit distance (also known as Levenstein distance). The Hamming distance between two strings is the minimal number of substitution operations required to transform the first string into the second string. While edit distance is the minimal number of substitution, insertion, or deletion operations required to transform the first string into the second string. On the other hand, wildcards matching problem is when the pattern has a wildcard, also known as don't care character, represented by $\Phi$, that can match with any other character in the alphabet set. For these matching problems, there is yet no linear (in structure size and query time) solution for them.

In this paper we are considering the following problems with the following inputs and outputs:

\textbf{Problem 1: Dictionary Matching} 

Inputs: 
	Given database of $N$ strings, each string $s$ is of length $m$ symbols of finite set of size $\Sigma$, and $\sum_{i=1}^{N} |s_{i}|  = n$; A pattern $P$ = $p_{1}p_{2} ...p_{m}$; and $K$. 
    
Outputs:
	All strings in the database that have with $P$ $\leq$ $K$ Hamming distance (K-Hamming distance), $\leq$ $K$ edit distance (K-edit distance), or $\leq$ $K$ wildcards matching (K-wildcards matching).

\textbf{Problem 2: Text Matching} 

Inputs: 
	Given a text of \textit{n} symbols of finite set of size $\Sigma$, a pattern $P$ = $p_{1}p_{2}...p_{m}$, and $K$.   
    
Outputs: All positions of the subsequences in the text that have with $P$ K-Hamming distance, K-edit distance, or K-wildcards matching.

Note that problem 1 is a simple case of problem 2. This paper will start describing the data structure for problem 1; as it is easier to be described than problem 2. After that, the design for problem 2 will be described. 

In this paper an algorithm is introduced for a novel tree structure that mainly leads to an efficient bound for the above problems.

\textit{Outline}: Section 2 surveys the background and related work of the problems. Then, section 3 states the preliminaries. Next, section 4 shows the design of the error tree structure for the first problem. After that, the modified design for problem 2 will be shown in section 5. Finally, section 6 states the conclusions.  

\section{Background and Related Work}

For the problem of text indexing, the naive algorithm takes $O(nm)$ time. A faster algorithm, Kangaroo method, was proposed by \cite{landau1986efficient} that build least common ancestor tree \cite{harel1984fast}, which allows to find each alignment in $O(k)$ time, hence the time cost will be $O(kn)$. A better algorithm was proposed by \cite{abrahamson1987generalized} that compute all alignments in $O(n\sqrt[]{mlogm})$. The recent algorithm takes $O(n\sqrt[]{klogk})$ which was proposed by \cite{Amir:2000:FAS:338219.338641}. Note that all the bounds above are in the order of $n$.  

Recently, \cite{cole2004dictionary} proposed a data structure that takes $O(n\frac{(c_{1}log n)^{k}}{k!})$ words and find the k-Hamming distance in $O(m + \frac{(c_{2}log n)^{k}loglogn}{k!} + occ)$ time, as well as a structure of $O(n\frac{(c_{3}log n)^{k}}{k!})$ words which finds the k-edit distance in $O(m + \frac{(c_{4}log n)^{k}loglogn}{k!} + 3^{k}occ)$, where $c_{1}, c_{2}, c_{3}, c_{4} > 1 $ are constants. The data structure of \cite{coelho2006dotted} can output the k-edit distance in $O(3^{k}m^{k+1} + occ)$. The construction of their index structure needs $O(nlog^{k}n)$ words of space in the average case, and takes $O(KN\Sigma)$ of time where $N$ is the number of nodes in the index. An upper bound algorithm was proposed by \cite{tsur2010fast}, where the space complexity of the index structure is $O(n^{1+\epsilon})$ for any constant $\epsilon > 0$, but with a query time of $O(m+loglogn+occ)$ for both k-Hamming distance and k-edit distance.   

Many algorithms solved both distances using a lower structure space but using an upper query time. Among these algorithms, \cite{chan2006linear} presented a linear space index $O(n)$, but with $O(m+log^{k(k+1)}nloglogn + occ)$ of query time. In \cite{lam2005improved}, a data structure of $O(n\sqrt[]{logn}log\Sigma)$ bits was proposed, that takes $O(\Sigma ^{k}m^{k}(k+loglogn)+occ)$ of query time; while using $O(n)$ bits of space, the query time will be $O(log^\epsilon n(\Sigma ^{k}m^{k}(k+loglogn)+occ))$, where $0 < \epsilon \leq 1$. Using more space, \cite{huynh2006approximate} showed an index structure that needs $O(nlogn)$ bits and takes $(\Sigma ^{k}m^{k} max(k,logn) + occ)$ of query time. They also reduced the index space to $O(n)$ by increasing the query time by factor of $O(logn)$.

For the k-wildcards matching problem, there is a light reduction in the time and space cost over the distances problems. The design of error tree structure in this paper allows solving the k-wildcards matching with the same bound of k-Hamming distance. Many algorithms proposed structures for solving the problem such as \cite{cole2004dictionary}, \cite{bille2014string}, and \cite{lewenstein2014less}. \cite{cole2004dictionary} proposed a structure of $O(n\frac{(k+log n)^{k}}{k!})$ words, that solves the problem in $O(m \frac{2^klog log n}{k!} + occ)$. While \cite{bille2014string} generalized the structure of \cite{cole2004dictionary} and reduced the space to $O(nlognlog_{\beta}^{k-1}n)$ words, but the query time increased to $O(m +  \beta^kloglogn +occ)$, where $2 \leq \beta \leq \Sigma$. Less space structure was presented in \cite{lewenstein2014less}, where the needed space is $O(nlog^{k}nlog\Sigma)$ bits, but with slight increase in the query time to $O(m +2^{k}logn +occ)$. 

Error Tree is a novel tree structure that is mainly oriented to solve the aforementioned problems. Firstly by constructing a tree structure that cost $O(n\frac{log_\Sigma ^{k}n}{k!})$ words  for Hamming distance and wildcards matching, and takes $O(\frac {m^k}{k!} + occ$)($O(m + \frac {log_\Sigma ^kn}{k!} + occ$) in the average case) of query time for any online/offline pattern. As well a tree structure of $O(2^{k}n\frac{log_\Sigma ^{k}n}{k!})$ words and $O(\frac {m^k}{k!} + 3^{k}occ$) ($O(m + \frac {log_\Sigma ^kn}{k!} + 3^{k}occ$ ) in the average case) of query time for edit distance of any online/offline pattern.

\section{Preliminaries}
For a string $s$, $len(s)$ is the length of string $s$, $s[x:y]$  is the substring from position $x$ to position $y$, and  $suff(s, i)$ is $ith$ suffix of string $s$. For a list $l$, $l[i]$ is the item at the index $i$, $l[-1]$ is the item at the last index, similarly $l[-2]$ is the item before the last item in $l$.

\section{Dictionary Matching Algorithm}
In order to explain the steps of the algorithm, the paper will firstly show the steps for k = 1, and then explain the general design when k $\geq$ 2.
\subsection{k = 1's Case}

The algorithm involves two stages; construction of a tree, then searching for the strings that have Hamming distance with $P$ equal to 1, a.k.a strings with 1-Hamming distance or 1-mismatch. 

\subsubsection{Construction stage}

\textbf{1.}	Firstly, a generalized suffix tree, \cite{DBLP:journals/algorithmica/Ukkonen95}, needs to be built for the strings in the database. So, by the definition of the suffix tree, we will have leaf node for each suffix of the strings in the database. $O(n)$ words.

\textbf{2.}	All leaves and internal nodes in the suffix tree will be assigned a unique key. $O(n)$ words.

\textbf{Definition 1}: A suffix tree that has a unique key that  identifies each leaf node and each internal node in the suffix tree is called \textit{keyed suffix tree (KST)}. 

\textbf{Definition 2}: For a string \textit{s} and a given KST, the function \textit{All Visited Nodes} denoted as \textit{AVN(s)} returns a list of the nodes$'$ keys and the edges$'$ lengths with their order that results from walking \textit{s} in the KST.

\textbf{Corollary 1}: For the case of k = 1 $and$ since in problem 1 we already have leaves for all suffixes of the strings in the database, we can find AVN($s$)[-1] for any suffix \textit{s} of any sting in the database in a constant time. Because AVN($s$)[-1] is the key of the last visited node after traversing the suffix $s$ in KST, which must be a leaf node. This can be by hashing all the leaves and returning AVN($s$)[-1] in a constant time without walking $s$ in the KST.

\textbf{Corollary 2}: Given a KST, and 2 strings, $s_{1}$ and $s_{2}$, where len($s_{1}$) = len($s_{2}$) and $s_{1}$ is in the KST. If Hamming distance ($s_{1}$, $s_{2}$) = 1, and the mismatch occur at position \textit{x} where \textit{x} is not the last position, then:
AVN(suff($s_{1}$, x+1)) = AVN(suff($s_{2}$, x+1)). Similarly AVN(suff($s_{1}$, x+1))[-1] = AVN(suff($s_{2}$, x+1))[-1].

\textbf{3.}	Construct a compact trie of all the strings in the database, $O(n)$ time and space.

\textbf{4.}	For each internal node $v$, a hash table ${I_{1}}$ is initialized. Then for all leaves $L$ of the subtree rooted at $v$, and assuming $v$ at level (symbol depth) $i$, then for all $l$ in $L$, we first pick any string $s$ labeled at $l$, then add to ${I_{1}}$ a tuple of $(AVN(suff(s, i+1)[-1], l)$. So:

\begin{verbatim}
For each internal node v in the tree:
     initialize a hash table I_1
     // get the level of the node
     i = get_level(v) 
     // get node desc   
     L = get_desc_leaves(v) 

     For l in L:
     	   pick a string s labeled at l
         v.I_1.add(AVN(suff(s, i+1)[-1], l) 
         
\end{verbatim}

\textbf{Definition 3}: We call such a tree structure, without loss of generality, a \textit{1-error tree} as it was constructed to find 1 mismatch.

Eventually, we will perform step 4 for $O(N)$ internal nodes, and at each node we will be bounded to the number of descendants leaves. If the 1-error tree is unbalanced, then we will not perform step 4 on the leaves under a branch that is on a heavy path (the path of most descendant leaves), namely at each node we have $\Sigma$ branches, all descendant leaves of the branch on a heavy path will be excluded in step 4 and will be treated in the query stage as edge. This means that balanced trie will be the worst case scenario. So, the bound will be \textit{O(Nlog$_{\Sigma}$N)} words of space.

\subsubsection{Query stage:}

Firstly, given the pattern \textit {P}, all its suffixes need to be added to the KST, and compute $AVN(.)[-1]$ for each suffix. So, the results will be the following list $R$:
    \{AVN(suff(P, 1))[-1], AVN(suff(P, 2)[-1]),..., AVN(suff(P, m))[-1]\}$_{m}$.
    
Secondly, we will walk $P$ in the 1-error tree as the following: If the walking is on edge, and the next symbol in $P$ match with the next symbol, then continue as exact match. If the next symbol in $P$ doesn't match with the next symbol at level $j$, this means that we reach 1 mismatch, and we can jump over the next symbol (since the walking is on edge) and continue as exact match until we reach a leaf, if any, outputs the strings labeled at that leaf as 1-Hamming distance at position $j$. 

Now, if the walking reaches a node $v$ where $v$ at level $j$, then look whether the key $R[j+1]$ is in the $I_{1}$ table of $v$ (constant time cost as $I_{1}$ is a hash table). If yes, all strings labeled at the leaves that were associated with key $R[j+1]$ have 1-Hamming distance at position $j$; then continue as exact match and the searching for k$\leq$1 mismatch. If next symbol in $P$ doesn't match with any child of $v$, then stop searching.

\subsubsection{Extension for indels}
The design can be extended to handle the operations of insertions and deletions; which means we can output all strings that have edit distance of $\leq$ K=1 with $P$, instead of only the Hamming distance.

Before of all, because insertion and deletion will cause shifting in the suffixes, such shifts must be tracked and manipulated by the design of the algorithm; mainly the AVN function. 

If two strings $s_{1}$ and $s_{2}$ have edit distance of score 1 caused by deletion operation at position $x$ of $s_{2}$, this means that $suff(s_{1}, x)[1:m-x-1] = suff(s_{2}, x+1)$. Now as AVN function starts at the root node and $must$ end up at a node that should has a \textit{unique key}, the design should guarantee that. For $suff(s_{2}, x+1)$, it must end up on a node and this will not cause any conflict in computing AVN. On the other hand, $suff(s_{1}, x)[1:m-x-1]$, which is actually 1=k level up of suffix $suff(s_{1}, x)$, may cause a conflict because it may not be a leaf node. Therefor, this position must be guaranteed to be a leaf node with a unique key. Thus, such a preprocessing step must be performed.

\textbf{1.} For each internal node $v$ in the 1-error tree, and for each leaf $l$ of all leaves $L$ of the subtree rooted at $v$, and assuming $v$ at level $i$. Firstly pick any string $s$ labeled at $l$, then we will walk up by 1=k level of the parent node of the leaf node of $suff(s, i)$. If we reach a node, let's say $x$, we will check if $x$ has a leaf node as a child, if not create new leaf node with unique key. If there is no node, then a new node with a unique key will be created, next as a child of this new node we will create a leaf node with a unique key. The cost will be $O(Nlog_{\Sigma}N)$ space and time. This will help to track the effects of shifting the suffixes because of the deletions and the insertions.

Insertions and deletions can occur in the pattern or in the strings. Before explaining the 4 cases, we will need to introduce the following corollary, $given$ the fact that step 1 was already performed:

\textbf{Corollary 3}: Given a KST and 2 strings, $s_{1}$ and $s_{2}$, where len($s_{1}$) = len($s_{2}$) = $m$  and $s_{1}$ is in the KST. If edit distance ($s_{1}$, $s_{2}$) = 1 and the edit operation is a deletion at position \textit{x} in $s_{2}$ where \textit{x} is not the last position, then:

AVN(suff($s_{1}$, x))[1:m-x-1] = AVN(suff($s_{2}$, x+1)). Similarly AVN(suff($s_{1}$, x))[-2] = AVN(suff($s_{2}$, x+1))[-1]. Note that suff($s_{1}$, x))[1:m-x-1] will always end up at a leaf node (that must has a unique key) because of step 1.

\textbf{Corollary 4}: Given a KST and 2 strings, $s_{1}$ and $s_{2}$, where len($s_{1}$) = len($s_{2}$) = $m$  and $s_{1}$ is in the KST. If edit distance ($s_{1}$, $s_{2}$) = 1 and the edit operation is an insertion at position \textit{x} in $s_{2}$ where \textit{x} is not the last position, then:

AVN(suff($s_{1}$, x+1)) = AVN(suff($s_{2}$, x)[1:m-x-1]). Note also that suff($s_{2}$, x)[1:m-x-1] will always end up at a leaf node (that must has unique key) because of step 1, so similarly AVN(suff($s_{1}$, x+1))[-1] = AVN(suff($s_{2}$, x))[-2].

\textbf{2.} For the edit distance, the following 4 cases will be possible:

\textbf{Case 1:} Deletion in the strings; Note that using the $I_{1}$ table we can handle this case. Based on corollary 3, we can check whether AVN(suff($P$, i)[-2] in $I_{1}$ or not. If yes, then all strings labeled at leaves that were associated with the key of AVN(suff($P$, i)[-2] must have edit distance with $P$ as a deletion in them at position $i$.

\textbf{Case 2:} Insertion in the strings; For this case, another hash table $I_{1\_ins}$ need to be initialized. Then step 4 of section 4.1.1 will be performed, but instead of adding (AVN(suff(s, $i+1$))[-1], l) into $I_{1}$, (AVN(suff(s, $i$))[-2], l) will be added to $I_{1\_ins}$. Note that because of step 1, AVN(suff(s, $i$))[-2] will always be a leaf node with a unique key. This will allow to check whether AVN(suff($P$, $i+1$)[-1] in $I_{1\_ins}$ or not, based on corollary 4. If yes, then all strings labeled at leaves that were associated with the key AVN(suff($P$, $i$)[-2] must have edit distance with $P$ as an insertion in them at position $i$.

Before proceeding to the next two cases, note that a deletion in the strings is similar to an insertion in the pattern. Likewise, an insertion in the strings is similar to a deletion in the pattern. 

\textbf{Case 3:} Deletion in the pattern; there is no need to modify the construction of the 1-error tree. This case can be computed by searching AVN(suff($P$, $i+1$)[-1] in $I_{1\_ins}$.

\textbf{Case 4:} Insertion in the pattern; this case can be computed by searching AVN(suff($P$, $i$)[-2] in $I_{1}$.

In conclusion, at each internal node we will have three tables correspondent to the operations of mismatch and insertion. The cost for step 1 will be $O(Nlog_{\Sigma}N)$ words of space and time. The cost for step 2, which is computing $I_{1\_ins}$ will be the same as computing \textit{$I_1$} in step 4 of section 4.1.1, which is $O(Nlog_{\Sigma}N)$ words of space.

\subsection{K $\geq$ 2's Case}
In the case of K=1, the main step in the design is that we associate the key of last node(leaf), AVN(.)[-1], of the suffixes with the leaves label. In the K $\geq$ 2 case, we will associate the keys of all the nodes that were returned by AVN(.) for a suffix  $s$, with all tuples, that have $s$, in the $I_{k-1}$ tables on the nodes on the path of $s$ in the (k-1)-error tree. Before describing the steps of the design, we will state the following corollary:  

\textbf{Corollary 5}: Given a KST and 2 strings, $s_{1}$ and $s_{2}$, where len($s_{1}$) = len($s_{2}$)  and $s_{1}$ is in the KST. If Hamming distance ($s_{1}$, $s_{2}$) = $k$ and the mismatches occur at positions $pos = \{p_{1}, p_{2}, ..., p_{k-1}, p_{k} \}$ and at each level of positions in $pos$ in the path to $s_{1}$ there is a node; then:

$AVN(s_{1}[1: p_{1} - 1])) = AVN(s_{2}[1: p_{1} - 1]))$, 

$AVN(s_{1}[p_{1} + 1: p_{2} - 1]))= AVN(s_{2}[p_{1} + 1: p_{2} - 1]))$,  

.

.

$AVN(s_{1}[p_{k-1} + 1: p_{k - 1}]))= AVN(s_{2}[p_{k-1} + 1: p_{k} - 1]))$

As well equivalently

$AVN(s_{1}[1: p_{1} - 1]))[-1] = AVN(s_{2}[1: p_{1} - 1]))[-1]$, 

$AVN(s_{1}[p_{1} + 1: p_{2} - 1]))[-1]= AVN(s_{2}[p_{1} + 1: p_{2} - 1]))[-1]$,  

.

.

$AVN(s_{1}[p_{k-1} + 1: p_{k - 1}]))[-1]= AVN(s_{2}[p_{k-1} + 1: p_{k} - 1]))[-1]$

\subsubsection{Construction stage}
\textbf{1.} The first step is to collect the nodes keys in KST. So, we will do the following for each internal node $v$:

\textbf{1.1 } At node $v$, we know the descendants leaves $L$ under that node and at what level assuming $i$. We firstly initialize a hash table $I_{k}$, then for each leave $l$ in $L$ we will pick any string $s$. Then, compute \textit{AVN(suff(s, i + 1))}; note that in the computation of AVN(), If the walking is at edge, then AVN() will return  the length of that edge and a tag indicating that we are at edge. If we walk at node, then it will return the key of that node and a tag indicating that we are at a node. Note that we will have $O(log_\Sigma n)$ items in $t$, since the balanced trie is the worst case scenario for the design.
 
\textbf{1.2 } After computing \textit{AVN(suff(s, i + 1))}, first walk in the k-error tree to the leaf $l$ with the skipping of 1 levels, and do the following:
\begin{verbatim}

if next node u in AVN(suff(s, i + 1)) is aligned in the middle of an edge: 
    
    v.I_k.add( ((u.key(), edge), l)) 

if next node u1 in AVN(suff(s, i + 1)) aligned with a node u2: 
    for each tuple p in u2.I_k-1 that has l:
         	v.I_k.add( (u1.key(), p[1]),..., p[k]))
    
\end{verbatim} 

Note that we will not walk explicitly to leaf $l$. We will check the alignment between a visited node in the path with a node in \textit{AVN(suff(s, i + 1))},  or the length of edge we visit in the path with edge's length in \textit{AVN(suff(s, i + 1))}, which is a simple convolution. So, we will have the following cases while walking to leaf $l$ in the k-error tree:

\textbf{Case 1:} Next node $u$ in \textit{AVN(suff(s, i + 1))} is aligned in the middle of edge in the k-error tree. Then, add to $I_{k}$ of $v$ a tuple contains the key of $u$, a tag indicating the alignment was at edge, and $l$. 

\textbf{Case 2:} Next node $u1$ in \textit{AVN(suff(s, i + 1))} is aligned to a node $u2$ while walking in the k-error tree. Then for each tuple $p$ that has $l$ in $I_{k-1}$ of $u2$; we will associate, as a tuple, the key of $u1$ and the items in $P$, in their order, then add the tuple into $I_{k}$ of $v$.

At the end of walking, note that each of a $O(log_\Sigma n)$ nodes' key may get associated/multiplied to a $O(\frac{log_\Sigma ^{k-1} n}{k!})$ tuples that are in $I_{k-1}$ during the walking to leaf $l$. So eventually, each level will cost $O(N \frac{log_\Sigma ^{k-1} n}{k!})$ words of space.  As we have $O(log_\Sigma N)$ levels, the bound will sum up to $O(N log_\Sigma N \frac{log_\Sigma ^{k-1} n}{k!})$.

\textbf{1.3} Steps 1.1 and 1.2 were performed for \textit{suff(s, i+1)} and on the tables $I_{k-1}$ and skipping 1 level. Similarly, we will do the same steps for \textit{suff(s, i+2)},...,\textit{suff(s, i+k-1)} on the tables of $I_{k-2}$,...,$I_{1}$ and skipping 2,..., k-1; respectively.

\textbf{2. }So far, we are covering the case where at each internal node, the symbols at the first $k-1$ levels are errors. But this will not cover the case where we will have all the first $k$ symbols after the internal nodes are actually errors. For this case, we need to perform step 4 of section 4.1.1, for suff(s, i + k) instead of suff(s, i + 1) suffixes as in the case of K = 1. The cost for this case will be as case k = 1, $O(N log_\Sigma N)$ words of space.

Eventually, this will lead us to have the cost of constructing k-error tree for any $k$ to be $O(N log_\Sigma N \frac{log_\Sigma ^{k-1} n}{k!})$ words.

\subsubsection{Query stage} When $k=1$, the number of possible error positions will be $m$, as $m$ is the length of $P$. When $k \geq 2$, the number of possible combinations of error positions would be ${m \choose k}$, which is bounded to $O(\frac {m^k}{k!})$.

Before stating the steps we need to describe the following cases:

\textbf{Case 1:} Walking a suffix in the KST will diverge at internal node. For this case, the design of the algorithm is already covering this case, as we have already marked all internal nodes in the KST, back in the error tree. 

\textbf{Case 2:} Walking a suffix in the KST will diverge in the middle of edge. In this case, we will allow jumping (skipping next symbol in the edge) $k$ times during the walking \textit{ at that edge (and/or any coming edge)}. If after $\leq$ k jumps, the walking ends up at a leaf; then deduct how many jumps were performed out of the $k$ mismatches value during the searching process. If the walking after $\leq$ k jumps ends at internal node, then this case is similar to case 1, but with deducting how many jumps were performed out of the $k$ mismatches value during the searching process. If after exactly k jumps we didn't reach a node (internal or leaf), this would mean that \textit{P will have no outputs at all of $k$ mismatches with any string in the database}, as one of its suffixes couldn't reach a leaf or an internal node after allowing k jumps (where jumps are representing mismatches). Note that counting the jumps is only at edges and not on any internal node, as the algorithm's design is already marking the internal nodes back in the error tree, and the jumps (assumingly errors) after these internal nodes are already accounted for in the design. 

As a result of these cases we will define the following function:

\textbf{Definition 4}: For a string \textit{s}, an integer $K$, and a given KST, the function \textit{All Visited Nodes with k jumps} denoted as \textit{AVNJ(s, k)} returns a list of the nodes$'$ keys and the edges$'$ lengths with their order that resulted from walking \textit{s} in the KST with allowing $k$ jumps (in case of mismatch) on only the edge; and the positions of the jumps, if occurred.

\textbf{1.} We will collect AVNJ(s, k) for each suffix $s_1, s_2,..., s_{m}$ of the pattern. After that, we will have $m$ lists. Let's call this list $R$.

\textbf{2.} Walk the pattern $P$ in the k-error tree, then at each internal node $v$, assuming $v$ is at level $i$. We will search if $I_{k}$ has any of the combinations of ${m-i \choose k}$ keys which can be extracted from the $R$ list. If so, report the leaves' labels that were associated with the keys combination as the output. If we walk on edge, we will perform skip (jump) over mismatches that we may reach in a simple convolution.

The overall bound for querying $k$ mismatches would be $O(\frac {m^k}{k!}) + occ)$ time. Note that in the average case computing AVN(.) for all suffixes would traverse $O(log_{\Sigma}n)$ nodes, therefore we get bound of $O(m + \frac {log_\Sigma ^kn}{k!} + occ)$.

\subsubsection{Extension for indels}
In order to handle the insertions and the deletions for any $k$, we will need to consider the following modifications: 

\textbf{1. }Guaranteeing that we have a leaf node at the $k$ level above all leaves in KST. For this, we will visit $k$ level above each leaf and perform step 1 in section 4.1.3. Note that during the construction of 1-error tree to (k-1)-error tree, we must have already created leaves nodes for each case of 1 to k-1.

\textbf{3. }For insertions: At each internal node, we will need to perform the steps in section 4.2.1 not for \textit{suff(s, i)[1:m-i-1]}; where $i$ is the level of the node, instead of \textit{suff(s, i)[1:m-i-k-1]}, then add the results into $I_{k\_ins}$ table.

\textbf{4. }Likewise, we need to perform step 2 of section 4.2.1, but for \textit{suff(s, i)[1:m-i-k]}.  

\textbf{5. }Note that the edit distance can be any combination of substitution, deletion, or insertion. For this, we will perform step 1.2, of section 4.2.1, for all the tables at node, namely $I_{k}$ and $I_{k\_ins}$ tables, not only $I_{k}$; then adding the results into a hash table $I_{k\_edit}$. This will add up an extra space of $2^{k}$ words. So, the total cost for building k-error tree that handle $k$ edit distance will be \textit{$O(2^{k} N log_\Sigma N \frac{log_\Sigma ^{k-1} n}{k!})$}). 

\textbf{6. }The number of combinations that will be needed to search for will increase in the factor of $3^{k}$, hence the query time for edit distance will be $O(\frac {m^k}{k!} + 3^{k}occ$)($O(m + \frac {log_\Sigma^kn}{k!} + 3^{k}occ$) in the average case).

\section{Algorithm Design for Text Indexing }
The design and the construction of the error tree for problem 2 are similar to the design and the construction of problem 1, but there are some differences. Here, we describe these differences and what preprocessing steps will be needed to resolve them in order to be able to apply the same design of problem 1.

\textbf{1.} The depth of all the paths in the suffix tree is not $\leq$ $m$. Paths with $> m$ depths are useless when we search for a pattern of length $m$. As well, this will add more costs in backward traversing of the tree and during the creation of new nodes. 

\textbf{2.} In problem 1, we have already leaves nodes for each suffix of the strings. In problem 2 this is not the case, because we have just one text string unlike problem 1. The leaves for the suffixes of each suffix \textit{(or specifically m-mer)} are not explicitly constructed.

Now in order to resolve these issues; we will perform the following:

\textbf{1.} All paths in the suffix tree must have depth $\leq m$. For this, we need to traverse all paths in the suffix tree and count the depth of the path by summing the lengths of the edges on each path. When depth = $m$ is reached, then if that point is already a node, trim all edges/nodes below that node, and store explicitly the labels of the descendants leaves. If that point is on edge, then create a leaf node at that point, after that copy and store explicitly the labels of the descendants leaves of its sink node, lastly trim the edge below that point. The cost of this step will be $O(n)$ time and space since we don't need to read the edges' symbols, instead we will just need to read the length of the edges (constant time), as well we will create $O(n)$ new nodes. We call such a suffix tree \textit{Trimmed Suffix Tree} of depth $m$, $TST_{m}$.

\textbf{2.} Starting from $TST_{m}$, we need to mark/tag suffixes of these suffixes similarly to the design of problem 1. Note that in problem 1 not all $m$ suffixes of the strings were considered in the design, since we only computed the AVN(.) for the suffixes of the descendants leaves under the internal nodes. Thus, $O(nlog _\Sigma n)$ of suffixes will be under consideration not $O(nm)$. 

Now in order to resolve this, note that after performing step 1 and for instance, the 6th suffix of suffix at position 1020 will be the prefix from root to position $m-6$ of the suffix $1026$. By this, the cost to guarantee/create a leaf node for the 6th suffix of suffix 1020 is to start from suffix 1026 leaf and walk as far as position $m-6$, then make sure we have a leaf node there or create a new one. Again, there is no need to walk on the edge explicitly to reach point $m-6$, as reading the length of the edges is enough. Hence, the cost will be $O(log_\Sigma n)$. In conclusion, guaranteeing/creating leaf node for each considered suffixes at the internal nodes will need an extra cost of $O(log_\Sigma n)$ time.

\textbf{3.} There is no need to build another compressed trie for the text in this problem. As we may consider the $TST_{m}$ as enough representation for all the $k$ error trees. So, all operations and all the k error trees can be constructed within the $TST_{m}$ or using independent trees.  

So, after making these modifications, we can build the k-error trees using the same steps in problem 1, and the cost will be  $O(n\frac{log_\Sigma ^{k}n}{k!})$ words. 

There is an exception for case k=1, because this case will need $O(n log_\Sigma ^{2}n)$ time but only $O(n log_\Sigma n)$ words of space; the reason is that we need to find the AVN()[-1] for the considered suffixes at each internal node in $O(log_\Sigma n)$ time, but we will just store the AVN()[-1] value which is of a constant space.  

\section{Conclusion}
In this paper, we introduce a tree structure that allows to solve non-trivial problems using very efficient bounds. For the problems of K-Hamming distance and K-wildcards matching, we propose a structure that needs $O(n\frac{log_\Sigma ^{k}n}{k!})$ words and takes $O(\frac {m^k}{k!} + occ$)($O(m + \frac {log_\Sigma ^kn}{k!} + occ$) in the average case) of query time for any online/offline pattern. A tree structure of $O(2^{k}n\frac{log_\Sigma ^{k}n}{k!})$ words and $O(\frac {m^k}{k!} + 3^{k}occ$)($O(m + \frac {log_\Sigma^kn}{k!} + 3^{k}occ$) in the average case) of query time for edit distance and any online/offline pattern.

\bibliographystyle{plain}
\bibliography{References}

\end{document}